\documentstyle[preprint,12pt,aps]{revtex}
\tightenlines

\begin{document}

\title{Analytic Solution of a Relativistic Two-dimensional Hydrogen-like Atom 
in a Constant Magnetic Field}

\author{V\'{\i}ctor M. Villalba\footnote{e-mail villalba@ivic.ivic.ve}}
\address{Centro de F\'{\i}sica\\
Instituto Venezolano de Investigaciones Cient\'{\i}ficas, IVIC\\
Apdo 21827, Caracas 1020-A, Venezuela}
\author{Ramiro Pino\footnote{e-mail rpino@ivic.ivic.ve}}
\address{Centro de Qu\'{\i}mica and Centro de F\'{\i}sica,\\
Instituto Venezolano de Investigaciones Cient\'{\i}ficas, IVIC\\
Apdo 21827, Caracas 1020-A, Venezuela}

\maketitle
\begin{abstract}
We obtain exact solutions of the Klein-Gordon and Pauli Schr\"odinger equations for 
a two-dimensional hydrogen-like atom in the presence of a constant magnetic field. 
Analytic solutions for the energy spectrum are obtained for
particular values of the magnetic field strength. The results are compared to
those obtained in the non-relativistic and spinless case. We obtain that the
relativistic spectrum does not present $s$ states. 
\end{abstract}

The study of the two-dimensional Hydrogen atom in the presence of
homogeneous magnetic fields has been a subject of active research during the
last years. Much work has been done in the framework of non-relativistic
quantum mechanics, computing the energy spectrum for different magnetic
field strengths. This problem is of practical interest in discussing single
and multiple-quantum well (superlattice) systems in semiconductor Physics.

The two-dimensional Schr\"odinger equation, with a Coulomb potential $-\frac{%
Z}{r}$ and a constant magnetic $\vec{{\cal B}}$ field, perpendicular to the
plane where the particle is, can be written in atomic units\footnote{%
In this paper we adopt the atomic units $\hbar=M=e=1$ in the CGS system.} as
follows 
\begin{equation}
\label{1}H\varphi =\frac 12(-i\nabla +\frac 12\vec{{\cal B}}\times \vec r%
)^2\varphi -\frac Zr\varphi =i\partial_t\varphi= E\varphi 
\end{equation}
Since we are dealing with a two-dimensional problem, we choose to work in
polar coordinates $(r,\vartheta ).$ The angular operator operator $%
-i\partial _\vartheta $ commutes with the Hamiltonian (\ref{1}),
consequently we can introduce the following ansatz for the eigenfunction 
\begin{equation}
\label{2}\varphi (\vec r)=\frac{\exp (im\vartheta )}{\sqrt{2\pi }}\frac{u(r)%
}{\sqrt{r}} 
\end{equation}
Substituting (\ref{2}) into (\ref{1}), we readily obtain that the radial
function $u(r)$ satisfies the second order differential equation

\begin{equation}
\label{ecua}\left[ -\frac 12\frac{d^2}{dr^2}+(m^2-\frac 14)\frac 1{r^2}+%
\frac{\omega _L^2r^2}2-\frac Zr+m\omega _L-E\right] u(r)=0 
\end{equation}
where $\omega _L={\cal B}/2c$ is the Larmor frequency, $E$ is the energy,
and $m$ the eigenvalue of the angular momentum. A general closed-form
solution to (\ref{ecua}) in terms of special functions does not exist\cite
{Bagrov} There are analytic expressions for the energy for particular values
of $\omega _L$ and $m$ as, pointed out by Lozanskii \cite{Lozanskii} and
more recently by Taut \cite{Taut1,Taut2}. The advantage of having exact
values of the energy spectrum for some values of the magnetic field strength
and the angular momentum becomes clear when we use numerical methods for
computing the energy for any value ${\cal B}$ and $m$, in particular for
higher excited states (Rydberg states) and high magnetic fields, as pointed
out by Taut \cite{Taut1}.

In this article we solve the two-dimensional hydrogen atom with an
homogeneous magnetic field (perpendicular to the plane in which the electron
is located) when relativistic corrections are considered. We obtain analytic
solutions of the 2D Klein-Gordon and we also compute the relativistic energy
spectrum for some particular values of the angular momentum $m$ and the
magnetic field strength ${\cal B}.$. Finally, we solve the nonrelativistic 
problem when spin corrections are included. 

The covariant generalization of the Klein-Gordon equation in the presence of
electromagnetic interactions takes the form \cite{Bagrov,Davydov} 
\begin{equation}
\label{3}\left( g^{\alpha \beta }(\nabla _\alpha -\frac icA_\alpha )(\nabla
_\beta -\frac icA_\beta )-c^2\right) \Psi =0
\end{equation}
where $g^{\alpha \beta }$ is the contravariant metric tensor, and $\nabla
_\alpha $ is the covariant derivative. Since we are working in a 2+1
spacetime, the metric tensor $g_{\alpha \beta }$ written in polar
coordinates $(t,r,\vartheta )$ takes the form: 
\begin{equation}
\label{4}g_{\alpha \beta }=diag(-1,1,r^2)
\end{equation}
and the vector potential $A^\alpha $ associated with a Coulomb and a
constant magnetic field interaction is 
\begin{equation}
\label{5}A^\alpha =(-\frac Zr,0,-\frac{{\cal B}r^2}2).
\end{equation}
With the help of the vector potential (\ref{5}), it is straightforward to
verify that the corresponding magnetic and electric fields satisfy the
invariant relations 
\begin{equation}
\label{seis}F_{\alpha \beta }F^{\alpha \beta }=2({\cal B}^2-{\cal E}^2)=2(%
{\cal B}^2-\frac{Z^2}{r^4})
\end{equation}
\begin{equation}
\label{siete}^{*}F_{\alpha \beta }F^{\alpha \beta }=0\ \rightarrow \vec {%
{\cal E}}\cdot \vec {{\cal B}}=0
\end{equation}

where $F^{\alpha \beta}$ is the electromagnetic field strength tensor, which
in a (2+1) spacetime has three independent components.

Expression (\ref{seis}) and (\ref{siete}) tell us that in fact $A^{\alpha}$ is
associated with a 2D Coulomb atom in a constant magnetic field perpendicular
to the plane where the particle is located. The corresponding $\vec {{\cal E}%
}$ and $\vec{{\cal B}}$ can be written in polar coordinates as follows: 
\begin{equation}
\label{E}\vec {{\cal E}}=-\frac Z{r^2}\hat e_r 
\end{equation}
\begin{equation}
\label{B}\vec {{\cal B}}={\cal B}\hat e_z. 
\end{equation}
Since the vector potential components do not depend on time or the angular
variable $\vartheta ,$ we have that the wave function $\Psi $, solution of
the Klein-Gordon equation (\ref{3}) can be written as

\begin{equation}
\label{6}\Psi (r,\vartheta ,t)=\frac{u(r)}{\sqrt{r}}\exp (im\vartheta -Et), 
\end{equation}
where the function $u(r)$ satisfies the second order ordinary differential
equation

\begin{equation}
\label{7}\frac{d^2u(r)}{dr^2}+\left( \frac{\frac 14-m^2+\frac{Z^2}{c^2}}{r^2}%
-\frac{m{\cal B}}c-c^2+\frac{E^2}{c^2}-\frac 14\frac{r^2{\cal B}^2}{c^2}+%
\frac{2EZ}{c^2r}\right) u(r)=0 
\end{equation}
Equation (\ref{7}) does not admit an exact solution in terms of special
functions, and therefore approximate methods have to be applied in order to
compute the energy spectrum. It is worth noting that the 2+1 relativistic
Coulomb problem can be solved in closed form \cite{Nieto} and the energy
levels are 
\begin{equation}
\label{8}E=c^2\left[ 1+\frac{Z^2}{c^2(n-\frac 12+\sqrt{m^2-\frac{Z^2}{c^2}}%
)^{2}}\right] ^{-1/2} 
\end{equation}
Also we have that the energy spectrum of a relativistic spinless particle in
a constant magnetic field satisfies the relation

\begin{equation}
\label{9}\frac{E^2}{c^2}-c^2=\frac{{\cal B}}c\left( 2n+m+\left| m\right|
+1\right) 
\end{equation}
We look for a series solution of eq. (\ref{7}). In order to do that, after
analyzing the asymptotic behavior of the solution $u(r)$ at $r=0$ and as $%
r\rightarrow \infty $, we write $u(r)$ in the form 
\begin{equation}
\label{10}u(r)=\exp (-\frac 14\frac{{\cal B}r^2}c)r^{(\frac 12+\sqrt{m^2-%
\frac{Z^2}{c^2}})}\sum_{n=0}a_nr^n
\end{equation}
substituting (\ref{10}) into (\ref{7}) and imposing that

\begin{equation}
\label{a0}a_0\neq 0 
\end{equation}
we obtain

\begin{equation}
\label{a1}a_1=-\frac{2EZ}{\left( 2\sqrt{m^2-\frac{Z^2}{c^2}}+1\right) c^2}a_0
\end{equation}
and for $n\geq 2$ we have 
\begin{eqnarray}
\label{papa}
\label{rec}\left[ n^2+2n\sqrt{m^2-\frac{Z^2}{c^2}}\right] a_n+2a_{n-1}\frac{EZ}{c^2}+ \nonumber \\
a_{n-2}\left[ \frac{E^2}{c^2}-c^2-\frac {\cal B}c(n-1+m+\sqrt{m^2-\frac{Z^2}{c^2}})\right] =0
\end{eqnarray}
Since only polynomial solutions of eq (\ref{7}) are bounded as $r\rightarrow
\infty $, the series given by the recurrence relation (\ref{rec}) terminates
at a certain $n$ when $a_n=0=a_{n+1}=0,$ $a_{n+i}=0$ for any positive
integer value of i. From the recurrence relation (\ref{rec}), we readily
obtain the following relation 
\begin{equation}
\label{ener}\frac{E^2}{c^2}-c^2=(m+\sqrt{m^2-\frac{Z^2}{c^2}}+n)\frac{{\cal B%
}}c
\end{equation}
for those values of the field strength ${\cal B}$ for which (\ref{10})
becomes a polynomial. After substituting (\ref{10}) into (\ref{7}), we
obtain a system of equations which gives the permitted values of ${\cal B}$.
It is worth mentioning that the relation (\ref{ener}) makes sense only when 
\begin{equation}
\label{ineq}m^2-\frac{Z^2}{c^2}>0
\end{equation}
a condition that forbids the existence of the $s$ energy levels ($m=0)$,
this is in fact a particularity of the relativistic Klein-Gordon solution,
which is not present in the standard Schr\"odinger framework.  Let us obtain
the first excited state of the relativistic Klein-Gordon 2+1 hydrogen atom.
In this particular case we have that only $a_0$ and $a_1$ are nonzero, and
the recurrence relations (\ref{a1}),(\ref{papa}) give 
\begin{equation}
\label{pri1}\frac{2EZ}ca_0+a_1(1+2\sqrt{m^2-\frac{Z^2}{c^2})}=0
\end{equation}
\begin{equation}
\label{pri2}a_0\frac {\cal B}{c}+\frac{2EZ}{c^2}a_1=0
\end{equation}
\begin{equation}
\label{pri3}\frac{E^2}{c^2}-c^2=\frac {\cal B}{c}(2+m+\sqrt{m^2-\frac{Z^2}{c^2})}
\end{equation}
from which we obtain that the energy is given by the expression 
\begin{equation}
\label{energ}E=c^2\left[ 1-\frac{4Z^2(2+m+\sqrt{m^2-\frac{Z^2}{c^2})}}{%
c^2(2\sqrt{m^2-\frac{Z^2}{c^2}}+1)}\right] ^{-1/2}
\end{equation}
for a magnetic field  ${\cal B}$ that can be obtained after substituting (%
\ref{energ}) into (\ref{ener}).
\begin{equation}
\label{magnetico}{\cal B} =\frac{4E^2Z^2}{c^3(1+2\sqrt{m^2-\frac{Z^2}{c^2}})}
\end{equation}
also, we have that for  n=3 ($a_2\neq 0,\ a_n=0,\ n>2)$ the corresponding
energy is given by the expression
\begin{equation}
\label{energ2}E=c^2\left[ 1-\frac{2Z^2(\sqrt{m^2-\frac{Z^2}{c^2}}+m+3)}{%
c^2(3+4\sqrt{m^2-\frac{Z^2}{c^2}})}\right] ^{-1/2}
\end{equation}
for ${\cal B}$ given by 
\begin{equation}
\label{magnetico2}{\cal B}=\frac{2E^2Z^2}{c^3}\left[ (3+4\sqrt{m^2-\frac{Z^2}{c^2}}%
\right] ^{-1}
\end{equation}

In order to obtain the magnetic field  for which analytic solutions of eq. (%
\ref{ecua}) are possible, we have to solve the system of equations given by
the recurrence relation (\ref{rec}). The following table shows all the
allowed magnetic fields for $2\leq n\leq 10$. Notice that for higher values
of n the number of solutions increases as int(n/2).

\begin{center}
{\scriptsize 
\begin{tabular}{|r|r|r|l|l|l|} \hline
\multicolumn{1}{|c|}{n} & \multicolumn{1}{|c|}{$\cal B$} &\multicolumn{1}{|c|}{$E$}
&\multicolumn{1}{|c|}{$E-c^2$} & \multicolumn{1}{|c|}{non rel. $E$} 
&\multicolumn{1}{|c|}{N} \\ \hline
2 &182.6958 &18770.3334 &1.3334  &1.3336 &1  \\ \hline 
3 &39.1452  &18769.4285 &0.4285  &0.4286 &2  \\ \hline
4 &15.1057  &18769.2205 &0.2205  &0.2205 &3  \\ \hline
4 &147.3011 &18771.1502 &2.1502  &2.1504 &2  \\ \hline
5 &7.4761   &18769.1364 &0.1364  &0.1364 &4  \\ \hline
5 &32.8207  &18769.5989 &0.5989  &0.5989 &3  \\ \hline
6 &4.2611   &18769.0933 &0.0933  &0.0933 &5  \\ \hline
6 &13.0381  &18769.2850 &0.2850  &0.2855 &4  \\ \hline
6 &125.4166 &18771.7461 &2.7461  &2.7464 &3  \\ \hline 
7 &2.6642   &18769.0686 &0.0686  &0.0686 &6  \\ \hline
7 &6.5938   &18769.1684 &0.1684  &0.1684 &5  \\ \hline
7 &28.5767  &18769.7300 &0.7300  &0.7300 &4  \\ \hline
8 &1.7784   &18769.0519 &0.0519  &0.0519 &5  \\ \hline
8 &3.8206   &18769.1115 &0.1115  &0.1115 &6  \\ \hline
8 &11.5616  &18769.3375 &0.3375  &0.3375 &5  \\ \hline
8 &110.2496 &18772.2186 &3.2186  &3.2190 &4  \\ \hline
9 &1.2462   &18769.0409 &0.0409  &0.0409 &8  \\ \hline
9 &2.4191   &18769.0794 &0.0794  &0.0794 &7  \\ \hline
9 &5.9334   &18769.1948 &0.1948  &0.1948 &6  \\ \hline
9 &25.4888  &18769.8372 &0.8372  &0.8372 &5  \\ \hline
10&0.9080   &18769.0331 &0.0331  &0.0331 &9  \\ \hline
10&1.6310   &18769.0595 &0.0595  &0.0595 &8  \\ \hline
10&3.4786   &18769.1269 &0.1269  &0.1269 &7  \\ \hline 
10&10.4433  &18769.3811 &0.3811  &0.3811 &6  \\ \hline
10&98.9813  &18772.6121 &3.6121  &3.6125 &5  \\ \hline 
\end{tabular}}

\medskip
\par

\small{{\bf Table 1} Energy values for all the allowed magnetic fields for $Z=1$, 
$m=-1$, and a comparison with the non relativistic energy spectrum. N is the number
of nodes of the radial wavefunction}

\end{center}

\medskip

A comparison with the non-relativistic energy levels, obtained with the help of 
the Schwartz method \cite{Schwartz}, which is a generalization of the mesh point 
technique for numerical approximation of functions, shows that the relativistic
correction to the problem becomes noticeable when $B>100$. Table 1  shows that the energy correction is smaller than 
$10^{-3}$ even when ${\cal B}\approx 100$.  From  (\ref{8}) and (\ref{9}) we can see that 
for weak as well as for strong magnetic field strengths, 
the r\^ole played by the relativistic corrections is to shift down the 
energy levels. 

The inclusion of spin effects can be carried out in a rather simple way with
the help of the Schr\"odinger-Pauli equation. 
\begin{equation}
\label{Pauli}H\Psi =\left[ \frac 12(\vec P+\frac{\vec A}c)^2+U(r)+\vec s\cdot \frac{\vec {\cal B}}c\right] \Psi 
\end{equation}
which, after substituting the vector potential (\ref{5}) into (\ref{Pauli}),
with $\Psi (\vec r)$ given by (\ref{6})  we obtain 
\begin{equation}
\label{Paulir}-\frac 12\frac{d^2u}{dr^2}+\left[ \frac 12(m^2-\frac 14)\frac 1%
{r^2}+\frac{\omega^2}{2}-\frac Zr\right] u=\left[ E-2\omega s -m%
\omega\right] u
\end{equation}
where ${\vec s}$ is the spin operator which satisfies the relations 
\begin{equation}
\vec s \cdot \vec s = 3/4, \hspace{1cm} s_i s_j = \frac{1}{4}(\delta_{ij}+i\epsilon_{ijk}s_k)
\end{equation}
and,  for our magnetic field strength, we have that ${\vec s} \cdot \vec{\cal B}$
can be written as 
\begin{equation}
\vec{s} \cdot \vec{\cal B} = \frac{1}{2} \sigma_3 {\cal B}
\end{equation}
where $\sigma_3$ is the diagonal Pauli matrix. 
Eq. (\ref{Paulir}) has essentially the same form of (\ref{ecua}). The
presence of the spin introduces a shift in the energy proportional to the
magnetic field strength. For the allowed magnetic field values reported in
\cite{Taut1} we have that the Schr\"odinger-Pauli spectrum is

\begin{equation}
\label{enerPauli}
E=\omega(n+2\sigma +m+\left| m\right| ) 
\end{equation}
It is worth noticing that the spin contribution does not introduce any 
restriction on the allowed values of $m$. It is the relativistic contribution
that forbids the existence of the $s$ values. The inclusion of spin correction 
as well as relativistic effects requires to deal with the 2+1 Dirac
equation in the background field given by (\ref{E}) and (\ref{B}). This problem will be discussed   
in a forthcoming publication. 

\acknowledgments We thank Dr. Juan Rivero for reading and improving the 
manuscript.

\end{document}